\documentclass[graphicx,twocolumn,jcp]{revtex4}
\usepackage{amssymb,mathtools}

\usepackage[varg]{txfonts}
\usepackage{color}
\usepackage{setspace}
\usepackage{relsize}

\usepackage{tikz}
\usepackage{filecontents,pgfplots}
\usetikzlibrary{calc,fadings,decorations.pathreplacing,hobby}
\usetikzlibrary{arrows,positioning}
\usetikzlibrary{patterns}
\usetikzlibrary{fit}
\usepackage{MnSymbol}

\pgfplotsset{select coords between index/.style 2 args={
    x filter/.code={
        \ifnum\coordindex<#1\fi
        \ifnum\coordindex>#2\fi
    }
}}

\usepgfplotslibrary{fillbetween}
\usetikzlibrary{shapes}
\tikzset{>=latex}

\newcommand{\fig}[1]{FIG.\,\ref{#1}}
\newcommand{\eq}[1]{Eq.\,\eqref{#1}}
\newcommand{\zit}[1]{Ref.\,\cite{#1}}

\newcommand{\sect}[1]{Sec.\,\ref{#1}}

\newcommand\orangeline[1][]{%
   \,\tikz[baseline]\draw[#1,very thick,orange](0,0.35*\ht\strutbox)--(1.5*\ht\strutbox,0.35*\ht\strutbox);}
\newcommand\blackline[1][]{%
   \,\tikz[baseline]\draw[#1,thick](0,0.35*\ht\strutbox)--(1.5*\ht\strutbox,0.35*\ht\strutbox);}

\begin{document}

\title{Self-assembly and entropic effects in pear-shaped colloid systems:\\ I. Shape sensitivity of bilayer phases in colloidal pear-shaped particle systems} 

\author{Philipp W. A. Sch\"onh\"ofer}
\email[]{Philipp.Schoenhoefer@fau.de}
\affiliation{College of Science, Health, Engineering and Education, Mathematics and Statistics, Murdoch University, 90 South Street, 6150 Murdoch, WA, Australia}
\affiliation{Institut f\"ur Theoretische Physik I, Friedrich-Alexander-Universit\"at Erlangen-N\"urnberg, Staudtstra\ss{}e 7, 91058 Erlangen, Germany}
\author{Matthieu Marechal}
\affiliation{Institut f\"ur Theoretische Physik I, Friedrich-Alexander-Universit\"at Erlangen-N\"urnberg, Staudtstra\ss{}e 7, 91058 Erlangen, Germany}
\author{Douglas J. Cleaver}
\affiliation{Materials and Engineering Research Institute, Sheffield Hallam University, Sheffield S1 1WB, UK}
\author{Gerd E. Schr\"oder-Turk}
\email[]{G.Schroeder-Turk@murdoch.edu.au}
\affiliation{College of Science, Health, Engineering and Education, Mathematics and Statistics, Murdoch University, 90 South Street, 6150 Murdoch, WA, Australia}
\affiliation{Department of Applied Mathematics, Research School of Physical Sciences and Engineering, The Australian National University, 0200 Canberra, ACT, Australia}
\affiliation{Department of Food Science, University of Copenhagen, Rolighedsvej 26, 1958 Frederiksberg C, Denmark}
\affiliation{Physical Chemistry, Center for Chemistry and Chemical Engineering, Lund University, Lund 22100, Sweden}
\date{\today}

\begin{abstract}
The role of particle shape in self-assembly processes is a double-edged sword. On the one hand, particle shape and particle elongation are often considered the most fundamental determinants of soft matter structure formation. On the other hand, structure formation is often highly sensitive to details of shape. Here we address the question of particle shape sensitivity for the self-assembly of hard pear-shaped particles, by studying two models for this system: a) the pear hard Gaussian overlap (PHGO) and b) hard pears of revolution (HPR) model. Hard pear-shaped particles, given by the PHGO model, are known to form a bicontinuous gyroid phase spontaneously. However, this model does not replicate an additive object perfectly and, hence, varies slightly in shape from a ''true'' pear-shape. Therefore, we investigate in the first part of this series the stability of the gyroid phase in pear-shaped particle systems. We show based on the HPR phase diagram that the gyroid phase does not form in pears with such ''true'' hard pear-shaped potential. Moreover, we acquire first indications from the HPR and PHGO pair-correlation functions that the formation of the gyroid is probably attributed to the small non-additive properties of the PHGO potential.
\end{abstract}

\maketitle 

In colloidal and soft matter science, the influence of particle shape on the geometry of self-assembled meso-structures and, hence, on their physical properties is well documented. To some approximation, colloids behave as hard particles that are subject to thermal Brownian motion. Similar to objects with hard-core potentials, they interact largely by volume exclusion effects, which are defined by their outline, and otherwise feel no energetic repulsion or attraction. The effect of shape is demonstrated, for instance, in dense collections of elongated nano- or microrods, which spontaneously develop a preferential particle direction and, consequently, introduce a distinguished global orientation, known as the nematic director \cite{VF1990,FMMcT1984}. Furthermore, it has been reported that the morphology of platonic and other polyhedral colloids can be used as a tool to create complex crystalline arrangements \cite{H-AEG2011,NGdeGvRD2012,DEG2011,DEG2012,GdeGvRD2013,DvANG2017,KCEG2018}. Hence, the manipulation of particle shapes is an auspicious mechanism to design self-assembled materials. However, the relationship between the shape of the constituent particles and the adopted self-assembled structure is not straightforward. While particle shape is beyond doubt an important determinant of structure formation, only a handful of quantifiable shape parameters could be related to long-ranged order directly. In colloidal self-assembly it is generally accepted that nematic order only occurs in particles that are sufficiently elongated, indicated by the aspect ratio between the length and width of the particle \cite{VF1990,BF1997,JJE2008,DEG2012,MDD2017}. Similarly, it has been shown that close-packed structures, like those based on the $\gamma$-brass lattice, require particles with a high isoperimetric quotient, which indicates the ratio between the particle's volume and its surface area \cite{DEG2012}.\\

In this article we focus on a related aspect, namely shape sensitivity upon self-assembly, which aggravates the prediction of collective behaviour in multi-particle systems by just the outline of the single constituents even further. Even if morphological parameters are identified necessary for the formation of certain mesostructures, the stability of these assemblies tend to be sensitive towards small changes in shape. The sensitivity to details of shape is presumably most clearly observed in hard-core systems. These systems are by design reduced to the shape of the inherent particles, which is defined by the hard interaction potentials. Already introducing a small degree of polydispersity into simple systems like the hard sphere fluid \footnote{The standard deviation of the diameter distribution has to be $\sigma_d\gtrapprox 0.08\bar{d}$ with the mean diameter $\bar{d}$}, can destabilise the crystalline into an amorphous phase for high densities \cite{BW1999}. Similarly in other hard particle mixtures, where depletion attractions between hard colloidal particles are induced by a solvent of surrounding small depletants, entropic forces are highly affected by the shape of colloids \cite{ADKF1998,BF2000,DPGAF2004,BDvR2012,AMV2016,GOT2018,GTMOWL2018} (for a more in-depth discussion about depletion see part 2 of this series \cite{SMCS-T2020_2}). The significant influence of shape becomes also apparent by comparing the phase behaviour of hard spherocylinders \cite{VF1990} and hard ellipsoids \cite{FMMcT1984} obtained by simulations. Even though the shapes of the individual particles seem similar, the smectic phase is only assembled by spherocylinders and not by ellipsoids.\\

Those observations are in accordance with other hard particle systems, which have been studied by investigating the intermediate stages of interpolations between two shapes. It has been shown, for example, that in systems of hard cubes, rounded edges have a significant influence on the cubical ordering of the crystalline phase \cite{BST2010,ZFvdLG2011,NGdeGvRD2012,RSAPPCDSI2015}. In addition to these superballs also various families of truncated polyhedra \cite{DEG2011,DEG2012,GdeGvRD2013,DvANG2017,KCEG2018}, elongated and twisted triangular prisms \cite{DD2016}, discs with adjustable thickness \cite{MD2010} and very recently also dimpled spheres with various dimple sizes \cite{WDvAG2019} have been studied. Here, it has been indicated that especially more complex particle arrangements are stable within a narrow window of shapes which makes them even more prone to small shape changes.\\ 

\begin{figure*}[t!]
\centering
\input{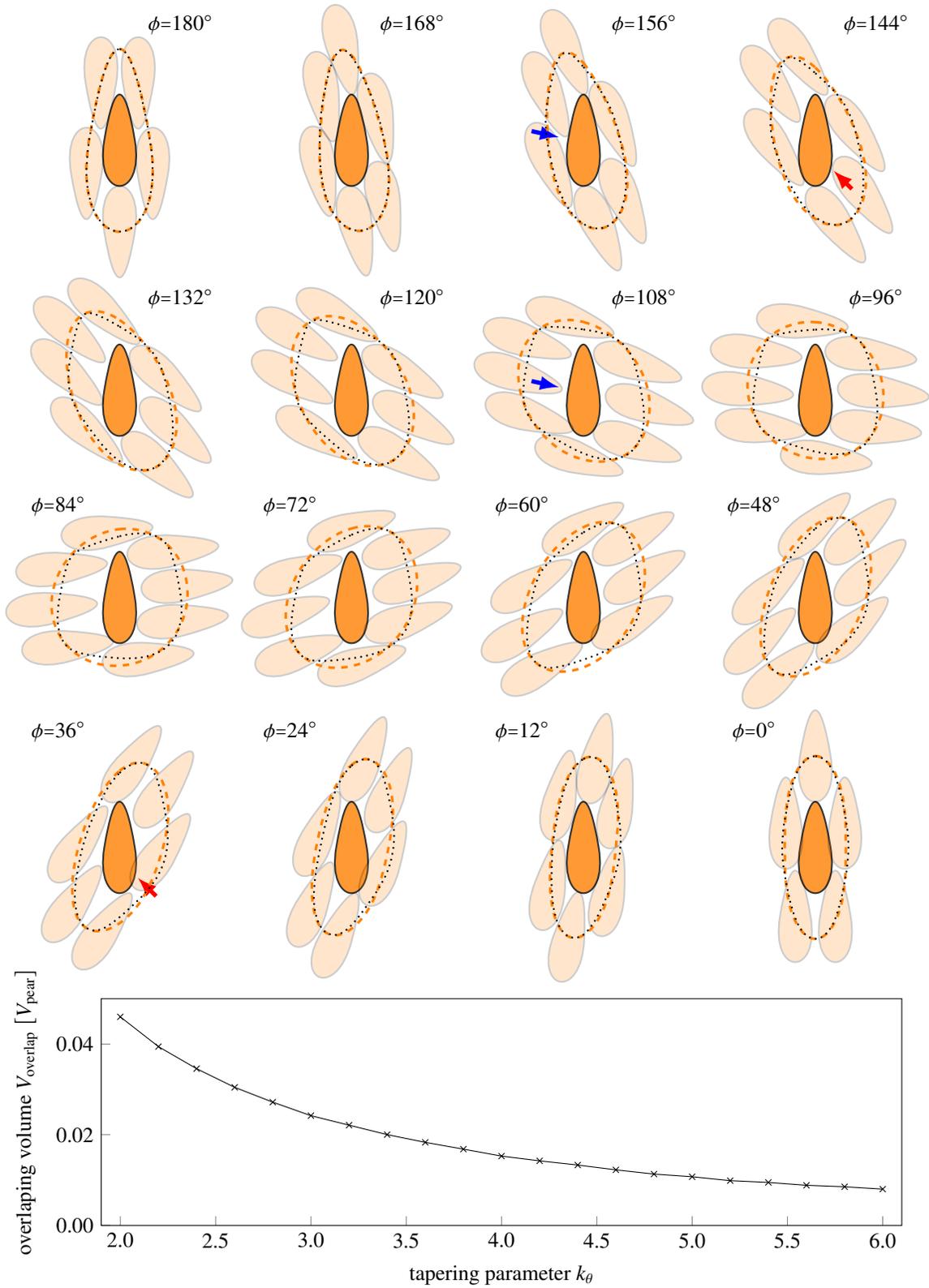}
\caption{Top: The contact profiles according to the PHGO model (\orangeline[dashed]) and the HPR model (\blackline[dotted]) for identical pear-shaped particles with $k=3$ and $\theta_k=15^\circ$ at different angles between the molecules $\phi=\arccos(\mathbf{u}_i{\cdot}\mathbf{u}_j)$ in the xz-plane. The surrounding pears are positioned in contact according to the PHGO model. The arrows showcase the different contact between blunt (red) and pointy (blue) ends depending on $\phi$. Bottom: The maximal overlap volume  $V_\text{overlap}$ between two PHGO particles with different tapering parameters $k_\theta$ when in contact. The volume is given in comparison to the volume of the B\'ezier pear $V_\text{pear}$.}
\label{fig:ContactFuntionPears}
\end{figure*}

Cubic structures based on triply periodic minimal surfaces are amongst the most complex representatives of such phases, which have been observed within the field of colloidal self-assembly. For instance, computational simulations of hard pear-shaped particles, reminiscent of tapered ellipsoids, indicate the spontaneous formation of highly symmetric liquid crystal phases, like the cubic and bicontinuous Ia$\bar{3}$d double gyroid \cite{EMBC2006,SEMCS-T2017} or the Pn$\bar{3}$m double diamond phase (upon addition of a hard sphere solvent) \cite{SCS-T2018}. Here, the shape of the used hard-core potential, called pear hard Gaussian overlap (PHGO) potential, is best illustrated by a pear shape, which is described by two B\'ezier curves \cite{BRZC2003} (see \fig{fig:ContactFuntionPears} for the outline of a pear-shaped particle). However, the effective shape of the PHGO model is just a close approximation and not a perfect fit to the B\'ezier description. Therefore, the shape represented by the PHGO potential can be interpreted as a slight distortion of the perfect B\'ezier pear.\\

Up to this point, the influence of the distinctions between the PHGO model and the ''true'' B\'ezier pear-shape has not been studied in detail, \textit{a fortiori}, as for the ellipsoidal counterparts (the hard Gaussian overlap (HGO) ellipsoids and the hard ellipsoids of revolution (HER) ) small differences between the two models are known \cite{P2008}. The phase transitions between the isotropic and orientationally ordered liquid crystal phases do not match perfectly for both ellipsoid models as the HGO interaction profile promotes the alignment of particles by a greater margin. Consequently, the phase transition of the HGO ellipsoids occurs for lower densities than for HER ellipsoids. Nevertheless, the distinct transition density does not change the characteristics of the observed phase behaviour significantly. Both models exhibit a similar nematic phase in between the isotropic and solid state without the HGO ellipsoids adding more complex phases. Thus, the two types of ellipsoids are qualitatively equivalent and their small differences in particle-shape are of only marginal consequences. However, the double gyroid phase is a much more complex structure than the ``simple'' nematic.\\

It seems plausible that higher complexity leads to an increased response and that especially the self-assembly of configurations like the double gyroid is more sensitive to the interaction of the particles. Hence, we focus in part 1 of this series on the phase behaviour of a more accurate, but computationally much more expensive B\'ezier pear model. In that case the hard potential is based on triangulated meshes of the pear-surface, which we address as the hard pears of revolution (HPR) model. Here, the contact is determined by testing for overlap between the triangulated surfaces and, hence, coincides with the B\'ezier description arbitrarily accurately.\\

In the following, we first detail the specific shape differences between the two pear-shaped particle models in \sect{sec:Micro}. Afterwards we analyse the effect of these distinctions by calculating the phase diagram of the HPR model numerically and comparing it to the phase behaviour of PHGO particles in \sect{sec:Meso}. Here we show that the gyroid phase, which can be interpreted as a warped bilayer phase, is not universal for tapered pear particles and that the special features of the PHGO contact function promote the formation of otherwise unfavourable bilayer-configurations. Subsequently in \sect{sec:Pair}, we analyse the local environment of the pear-shaped particles within the different phases. In combination with our results from part 2, where we observe the depletion behaviour between pear-shaped particles within a hard sphere solvent \cite{SMCS-T2020_2}, this study sheds light on the different mesoscopic behaviour between the PHGO and HPR model from a microscopic perspective.\\

\section{Microscopic differences between hard pears of revolution and pear hard Gaussian overlap particles}
\label{sec:Micro}

In \fig{fig:ContactFuntionPears} the contact profiles of PHGO and HPR particles with aspect ratio $k=3$ and tapering parameter $k_\theta=3$ are compared. The contact profile is determined by the interface of the excluded volume given by the contact function
\begin{equation}
\sigma(\mathbf{r}_{ij},\mathbf{u}_i,\mathbf{u}_j) =
\begin{cases}
0, \text{if particle } i \text{ and } j \text{ do not overlap},\\
1, \text{if particle } i \text{ and } j \text{ overlap}\\
\end{cases}
\end{equation}
with the relative distance $\mathbf{r}_{ij}$ between the reference particle $i$ and a secondary particle $j$ and their orientation vectors $\mathbf{u}_i$ and $\mathbf{u}_j$. It becomes apparent that the two models show considerable differences for relative angles $\phi=\arccos(\mathbf{u}_i{\cdot}\mathbf{u}_j)$ between $50^{\circ}$ and $130^{\circ}$. In this regime the PHGO profile often overestimates the overlap, which leads to gaps between the particles. This, however, is inherited from a similar error between the HGO and HER (hard ellipsoids of revolution) potential of the ellipsoid \cite{P2008}. For small angles an additional effect occurs. At around $30^{\circ}$ the PHGO profile also occasionally underestimates the contact distance, in other words the distance of closest approach, $\sigma$ compared to the B\'ezier shape such that the colloidal particles overlap with their blunt ends when represented by B\'ezier pears. The gap size and the overlap volume (see \fig{fig:ContactFuntionPears}) are higher for more asymmetrical pears, such that the PHGO approximation is worse for B\'ezier-pears with larger taper.\\

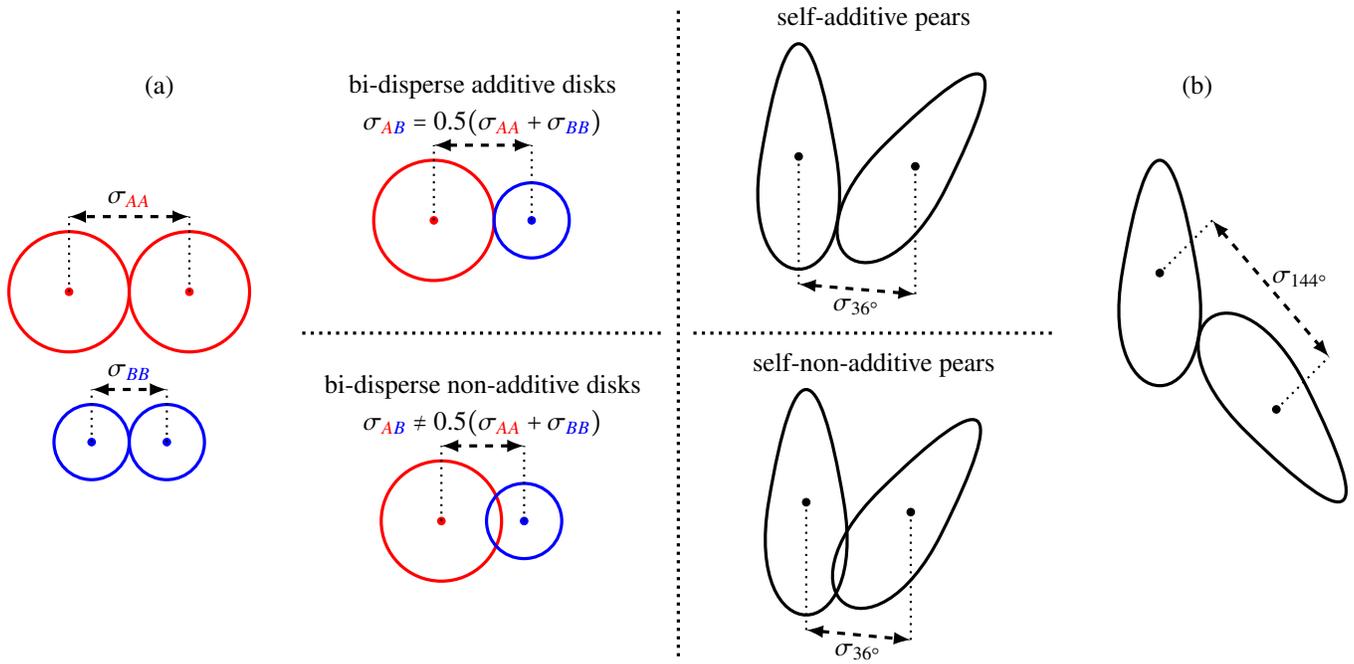
\begin{figure*}[t!]
\centering
\begin{tikzpicture}

    \node[anchor=south] at (0,3) {(a)};
    \node[anchor=south] at (13.8,3) {(b)};

    \begin{scope}[shift={(-0.4,0.55)}]
	\draw[red,very thick] (0.8,0) circle (0.8);	
	\draw[fill=red,red] (0.8,0) circle (0.05);	
	\draw[red,very thick] (-0.8,0) circle (0.8);	
	\draw[fill=red,red] (-0.8,0) circle (0.05);	
	\draw[very thick,dashed,<->] (0.8,1) -- (-0.8,1);	
	\draw[thick,dotted] (0.8,0) -- (0.8,1);	
	\draw[thick,dotted] (-0.8,0) -- (-0.8,1);	
    \node[anchor=south] at (0,1) {$\sigma_{\textcolor{red}{AA}}$};
    \end{scope}

    \begin{scope}[shift={(-0.4,-1.45)}]
	\draw[blue,very thick] (-0.5,0) circle (0.5);	
	\draw[fill=blue,blue] (-0.5,0) circle (0.05);	
	\draw[blue,very thick] (0.5,0) circle (0.5);	
	\draw[fill=blue,blue] (0.5,0) circle (0.05);	
	\draw[very thick,dashed,<->] (-0.5,0.7) -- (0.5,0.7);	
	\draw[thick,dotted] (0.5,0) -- (0.5,0.7);	
	\draw[thick,dotted] (-0.5,0) -- (-0.5,0.7);	
    \node[anchor=south] at (0,0.7) {$\sigma_{\textcolor{blue}{BB}}$};
    \end{scope}

	\draw[very thick,dotted] (6.9,-4.3) -- (6.9,4.3);	
	\draw[very thick,dotted] (1.9,0) -- (6.7,0);	
	\draw[very thick,dotted] (7.1,0) -- (11.9,0);	

    \begin{scope}[shift={(4.3,1.5)}]
	\draw[red,very thick] (-0.65,-0) circle (0.8);	
	\draw[fill=red,red] (-0.65,0) circle (0.05);	
	\draw[blue,very thick] (0.65,-0) circle (0.5);	
	\draw[fill=blue,blue] (0.65,0) circle (0.05);	
	\draw[very thick,dashed,<->] (-0.65,1) -- (0.65,1);	
	\draw[thick,dotted] (0.65,0) -- (0.65,1);	
	\draw[thick,dotted] (-0.65,0) -- (-0.65,1);	
    \node[anchor=south] at (0,1) {$\sigma_{\textcolor{red}{A}\textcolor{blue}{B}}=0.5(\sigma_{\textcolor{red}{AA}}+\sigma_{\textcolor{blue}{BB}})$};
    \node[anchor=south] at (0,1.5) {bi-disperse additive disks};
    \end{scope}

    \begin{scope}[shift={(4.3,-2.5)}]
	\draw[red,very thick] (-0.55,0) circle (0.8);	
	\draw[fill=red,red] (-0.55,0) circle (0.05);	
	\draw[blue,very thick] (0.55,0) circle (0.5);	
	\draw[fill=blue,blue] (0.55,0) circle (0.05);	
	\draw[very thick,dashed,<->] (0.55,1) -- (-0.55,1);	
	\draw[thick,dotted] (0.55,0) -- (0.55,1);	
	\draw[thick,dotted] (-0.55,0) -- (-0.55,1);	
    \node[anchor=south] at (0,1) {$\sigma_{\textcolor{red}{A}\textcolor{blue}{B}}\neq0.5(\sigma_{\textcolor{red}{AA}}+\sigma_{\textcolor{blue}{BB}})$};
    \node[anchor=south] at (0,1.5) {bi-disperse non-additive disks};
    \end{scope}

\begin{scope}[shift={(13.3,0.8)},scale=1]
	\draw[very thick] (-0.5,0) .. controls (-0.166666666,2) and (0.166666666,2) .. (0.5,0) .. controls (0.833333333,-2) and (-0.8333333333,-2) .. (-0.5,0) ;
	\draw[fill=black] (0,0) circle (0.05);	
	\draw[very thick, shift={(1.55,-1.81506)},rotate=-144] (-0.5,0) .. controls (-0.166666666,2) and (0.166666666,2) .. (0.5,0) .. controls (0.833333333,-2) and (-0.8333333333,-2) .. (-0.5,0) ;
	\draw[fill=black] (1.55,-1.81506) circle (0.05);	
	\draw[thick,dotted] (0.7,0.7) -- (0,0);	
	\draw[thick,dotted] (1.55,-1.81506) -- (2.25,-1.11506);	
	\draw[very thick,dashed,<->] (0.7,0.7) -- (2.25,-1.11506);	
    \node[anchor=south west] at (1.375,-0.30753) {$\sigma_{144^{\circ}}$};
    \end{scope}
    
\begin{scope}[shift={(8.5,2.35)},scale=1]
    \draw[very thick] (-0.5,0) .. controls (-0.166666666,2) and (0.166666666,2) .. (0.5,0) .. controls (0.833333333,-2) and (-0.8333333333,-2) .. (-0.5,0) ;
	\draw[fill=black] (0,0) circle (0.05);	
    \draw[very thick, shift={(1.55,-0.131359)},rotate=-36] (-0.5,0) .. controls (-0.166666666,2) and (0.166666666,2) .. (0.5,0) .. controls (0.833333333,-2) and (-0.8333333333,-2) .. (-0.5,0) ;
	\draw[fill=black] (1.55,-0.131359) circle (0.05);	
	\draw[thick,dotted] (0,-1.7) -- (0,0);	
	\draw[thick,dotted] (1.55,-0.131359) -- (1.55,-1.831359);	
	\draw[very thick,dashed,<->] (0,-1.7) -- (1.55,-1.831359);	
    \node[anchor=north] at (0.775,-1.7656795) {$\sigma_{36^{\circ}}$};
    \node[anchor=south] at (1,1.55) {self-additive pears};

\end{scope}

\begin{scope}[shift={(8.6,-2.25)},scale=1]
    \draw[very thick] (-0.5,0) .. controls (-0.166666666,2) and (0.166666666,2) .. (0.5,0) .. controls (0.833333333,-2) and (-0.8333333333,-2) .. (-0.5,0) ;
	\draw[fill=black] (0,0) circle (0.05);	
    \draw[very thick, shift={(1.38963,-0.131359)},rotate=-36] (-0.5,0) .. controls (-0.166666666,2) and (0.166666666,2) .. (0.5,0) .. controls (0.833333333,-2) and (-0.8333333333,-2) .. (-0.5,0) ;
	\draw[fill=black] (1.38963,-0.131359) circle (0.05);	
	\draw[thick,dotted] (0,-1.7) -- (0,0);	
	\draw[thick,dotted] (1.38963,-0.131359) -- (1.38963,-1.831359);	
	\draw[very thick,dashed,<->] (0,-1.7) -- (1.38963,-1.831359);	
    \node[anchor=north] at (0.694815,-1.7656795) {$\sigma_{36^{\circ}}$};
    \node[anchor=south] at (0.9,1.55) {self-non-additive pears};
\end{scope}

\end{tikzpicture}
\caption{a) The concept of an additive and non-additive mixture of disc species $A$ and $B$. In the additive mixture the interspecies contact distance $\sigma_{AB}$ can be calculated from the contact between disks of the same species $\sigma_{AA}$ and $\sigma_{BB}$ by an additive rule. In the non-additive case this rule does not hold. b) The concept of self-additive and self-non-additive system by the example of pear-shaped particles. The contact between different parts of self-additive pears at a certain relative angle (i.e $\phi=36^{\circ}$) and distance can be deduced logically from the contact between the same particles at a different angle (i.e $\phi=144^{\circ}$). In self-non-additive systems the contact distance between parts of the particles vary and do not follow an overall shape.}
\label{fig:additive}
\end{figure*}

In the following, we will use the term \textit{self-non-additivity} to describe this combination between over- and underestimation of the contact distance and this special angle dependency of the contact distance. Conventionally, hard-core interactions are labelled additive, if in a mixture the distance of closest approach $\sigma_{AB}$ between species $A$ and $B$ can be logically deduced from the contact distance between particles of the same type by the additive constraint: $\sigma_{AB}= 0.5(\sigma_{AA}+\sigma_{BB})$. If this rule does not hold, the mixture is referred to as non-additive \cite{BPGG1986,LALA1996,RE2001,HS2010,ZFLSSO'H2015}. This concept is illustrated in \fig{fig:additive}a.\\

A similar effect, however, also occurs in the mono-disperse PHGO particle system. This becomes apparent by explaining the choice of the \mbox{prefix} ``self'' in self-non-additivity which is illustrated by analysing the contact distance between the blunt ends of the pear-shaped particles in \fig{fig:ContactFuntionPears} and explained additionally in \fig{fig:additive}b. For certain relative angles, the blunt ends overlap ($\phi=36^{\circ}$), whereas for other angles their contact coincides with the B\'ezier description ($\phi=144^{\circ}$; indicated by red arrows in \fig{fig:ContactFuntionPears}). Similar behaviour is observed for the contact between the thin ends (gaps at $\phi=108^{\circ}$ and no gap at $\phi=156^{\circ}$; indicated by blue arrows in  \fig{fig:ContactFuntionPears}). Hence, the PHGO model represents the hard interactions between two B\'ezier pear-shaped object depending on their relative angle differently well. Alternatively, differently orientated pears can be interpreted as distinct hard particle species with non-additive interactions as the contact at $\phi=36^{\circ}$ can not be deduced additively form the contact at $\phi=144^{\circ}$ (see \fig{fig:additive}b). Moreover, the described angular dependency of the contact function implies that a true physical hard shape cannot copy the PHGO model \footnote{Additional overlap rules (like adding non-additive features to the blunt ends) are required to imitate the interactions between PHGO particles with physical hard shapes.}.\\

Evidently, the self-non-additivity of the PHGO model is a specific form of an orientation- and distance-dependent interaction potential. The interaction remains, for all relative orientations of the particles, a hard-core interaction where the particles experience no interaction until the point of contact.\\

\section{Phase hehaviour of hard pears of revolution and pear hard Gaussian overlap particles}
\label{sec:Meso}

The key result of this paper is the computation of the phase diagram of HPR particles and its comparison to the phase behaviour of pears as approximated by the PHGO model. Whereas PHGO particles were found to form complex phases (including smectic and gyroid), these phases are absent in the phase diagram of hard pears of revolution (HPR).\\

\subsection{Phase behaviour of pear hard Gaussian overlap (PHGO) particles}

\begin{figure*}[t]
\centering
\input{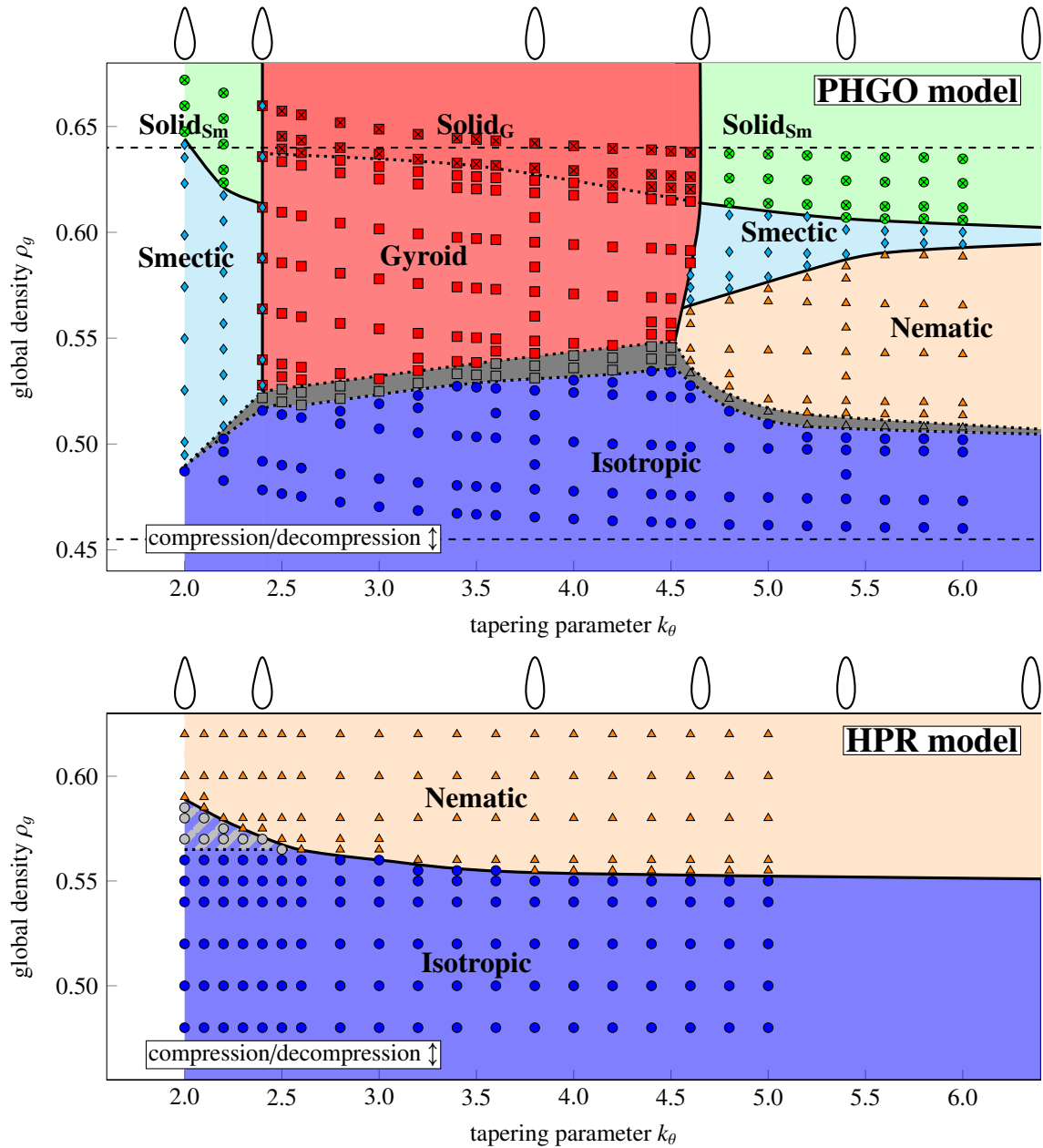}
\begin{tikzpicture}[
hatch distance/.store in=\hatchdistance,
hatch distance=10pt,
hatch thickness/.store in=\hatchthickness,
hatch thickness=2pt
]
\makeatletter
\pgfdeclarepatternformonly[\hatchdistance,\hatchthickness]{flexible hatch}
{\pgfqpoint{0pt}{0pt}}
{\pgfqpoint{\hatchdistance}{\hatchdistance}}
{\pgfpoint{\hatchdistance-1pt}{\hatchdistance-1pt}}%
{
\pgfsetcolor{\tikz@pattern@color}
\pgfsetlinewidth{\hatchthickness}
\pgfpathmoveto{\pgfqpoint{0pt}{0pt}}
\pgfpathlineto{\pgfqpoint{\hatchdistance}{\hatchdistance}}
\pgfusepath{stroke}
}
    \draw[thick,shift={(1.14cm,5.8cm)},scale=0.25] (-0.5,0) .. controls (0,2) and (0,2) .. (0.5,0) .. controls (1,-2) and (-1,-2) .. (-0.5,0) ;
    \draw[thick,shift={(2.27cm,5.8cm)},scale=0.25] (-0.5,0) .. controls (-0.083333333,2) and (0.083333333,2) .. (0.5,0) .. controls (0.916666666,-2) and (-0.916666666,-2) .. (-0.5,0) ;
    \draw[thick,shift={(6.25cm,5.8cm)},scale=0.25] (-0.5,0) .. controls (-0.236842105,2) and (0.236842105,2) .. (0.5,0) .. controls (0.763157894,-2) and (-0.763157894,-2) .. (-0.5,0) ;
    \draw[thick,shift={(8.665cm,5.8cm)},scale=0.25] (-0.5,0) .. controls (-0.282608695,2) and (0.282608695,2) .. (0.5,0) .. controls (0.717391304,-2) and (-0.717391304,-2) .. (-0.5,0) ;         
    \draw[thick,shift={(10.79cm,5.8cm)},scale=0.25] (-0.5,0) .. controls (-0.314814814,2) and (0.314814814,2) .. (0.5,0) .. controls (0.685185185,-2) and (-0.685185185,-2) .. (-0.5,0) ;        
    \draw[thick,shift={(13.49cm,5.8cm)},scale=0.25] (-0.5,0) .. controls (-0.346153846,2) and (0.346153846,2) .. (0.5,0) .. controls (0.653846153,-2) and (-0.653846153,-2) .. (-0.5,0) ; 
    \begin{axis}[
        xlabel=tapering parameter $k_{\theta}$,
        ylabel=global density $\rho_g$,
        xtick pos=left,
        ytick pos=left,
        xmin = 1.6,
        xmax=6.4,
        ymin = 0.455,
        ymax = 0.63,
        width=152.1mm,
        height=69.375mm,
        xtick = {2.0,2.5,3.0,3.5,4.0,4.5,5.0,5.5,6.0},
        ytick = {0.50,0.55,0.60},
        x tick label style={
        /pgf/number format/.cd,
            fixed,
            fixed zerofill,
            precision=1,
        /tikz/.cd
        },
        y tick label style={
        /pgf/number format/.cd,
            fixed,
            fixed zerofill,
            precision=2,
        /tikz/.cd
    }
        ]
              
      \addplot[name path=i1,smooth,black,very thick] plot coordinates {
        (2,0.589)
        (2.2,0.579)
        (2.4,0.571)
        (2.6,0.5648)
        (3,0.56)
        (3.8,0.554)
        (6.4,0.551)
    };
     \addplot[name path=i2,smooth,black,very thick,dotted] plot coordinates {
        (2,0.565)
        (2.55,0.565)
    };    
    \addplot[name path=daxis1,draw=none, mark=none,domain=2.0:6.5] {0.45};  
     \addplot[name path=taxis1,draw=none, mark=none,domain=2:3] {0.63};
     \addplot[name path=taxis2,draw=none, mark=none,domain=2:6.5] {0.63};

     \addplot[thick,mark=none,domain=1.5:6.5] {0.63};
     \addplot[thick,mark=none,domain=1.5:6.5] {0.455};
     
    \addplot[blue,opacity=0.5] fill between [ of=i1 and daxis1];
    \addplot[pattern=flexible hatch,pattern color=lightgray,hatch thickness=3pt] fill between [ of=i2 and taxis1];
    \addplot[white] fill between [ of=i1 and taxis2];
    \addplot[orange,opacity=0.2] fill between [ of=i1 and taxis2];

     \addplot[every mark/.append style={solid, fill=blue},mark=*,draw=none] coordinates {
     (2.0,0.48)
     (2.1,0.48)
     (2.2,0.48)
     (2.3,0.48)
     (2.4,0.48)
     (2.5,0.48)
     (2.6,0.48)
     (2.8,0.48)
     (3.0,0.48)
     (3.2,0.48)
     (3.4,0.48)
     (3.6,0.48)
     (3.8,0.48)
     (4.0,0.48)
     (4.2,0.48)
     (4.4,0.48)
     (4.6,0.48)
     (4.8,0.48)
     (5.0,0.48)

     (2.0,0.5)
     (2.1,0.5)
     (2.2,0.5)
     (2.3,0.5)
     (2.4,0.5)
     (2.5,0.5)
     (2.6,0.5)
     (2.8,0.5)
     (3.0,0.5)
     (3.2,0.5)
     (3.4,0.5)
     (3.6,0.5)
     (3.8,0.5)
     (4.0,0.5)
     (4.2,0.5)
     (4.4,0.5)
     (4.6,0.5)
     (4.8,0.5)
     (5.0,0.5)
     
     (2.0,0.52)
     (2.1,0.52)
     (2.2,0.52)
     (2.3,0.52)
     (2.4,0.52)
     (2.5,0.52)
     (2.6,0.52)
     (2.8,0.52)
     (3.0,0.52)
     (3.2,0.52)
     (3.4,0.52)
     (3.6,0.52)
     (3.8,0.52)
     (4.0,0.52)
     (4.2,0.52)
     (4.4,0.52)
     (4.6,0.52)
     (4.8,0.52)
     (5.0,0.52)
     
     (2.0,0.54)
     (2.1,0.54)
     (2.2,0.54)
     (2.3,0.54)
     (2.4,0.54)
     (2.5,0.54)
     (2.6,0.54)
     (2.8,0.54)
     (3.0,0.54)
     (3.2,0.54)
     (3.4,0.54)
     (3.6,0.54)
     (3.8,0.54)
     (4.0,0.54)
     (4.2,0.54)
     (4.4,0.54)
     (4.6,0.54)
     (4.8,0.54)
     (5.0,0.54)
     
     (2.0,0.55)
     (2.1,0.55)
     (2.2,0.55)
     (2.3,0.55)
     (2.4,0.55)
     (2.5,0.55)
     (2.6,0.55)
     (2.8,0.55)
     (3.0,0.55)
     (3.2,0.55)
     (3.4,0.55)
     (3.6,0.55)
     (3.8,0.55)
     (4.0,0.55)
     (4.2,0.55)
     (4.4,0.55)
     (4.6,0.55)
     (4.8,0.55)
     (5,0.55)
     
     (3.4,0.555)
     (3.2,0.555)
     (3.6,0.555)
     
     (2.0,0.56)
     (2.1,0.56)
     (2.2,0.56)
     (2.3,0.56)
     (2.4,0.56)
     (2.5,0.56)
     (2.6,0.56)
     (2.8,0.56)
     (3,0.56)

     };
     
     \addplot[every mark/.append style={solid, fill=lightgray},mark=*,draw=none] coordinates {

     (2.5,0.565)
     
     (2.0,0.57)
     (2.1,0.57)
     (2.2,0.57)
     (2.3,0.57)
     (2.4,0.57)

     (2.2,0.575)

     (2.0,0.58)
     (2.1,0.58)

     (2.0,0.585)
    };

     \addplot[every mark/.append style={solid, fill=orange},mark=triangle*,draw=none] coordinates {

     (3.8,0.555)
     (4,0.555)
     (4.2,0.555)
     (4.4,0.555)
     (4.6,0.555)
     (4.8,0.555)
     (5,0.555)

     (3.2,0.56)
     (3.4,0.56)
     (3.6,0.56)
     (3.8,0.56)
     (4,0.56)
     (4.2,0.56)
     (4.4,0.56)
     (4.6,0.56)
     (4.8,0.56)
     (5,0.56)

     (2.6,0.565)
     (2.8,0.565)
     (3,0.565)

     (2.5,0.57)
     (2.6,0.57)
     (2.8,0.57)
     (3,0.57)

     (2.3,0.575)
     (2.4,0.575)

     (2.2,0.58)
     (2.3,0.58)
     (2.4,0.58)
     (2.5,0.58)
     (2.6,0.58)
     (2.8,0.58)
     (3,0.58)
     (3.2,0.58)
     (3.4,0.58)
     (3.6,0.58)
     (3.8,0.58)
     (4,0.58)
     (4.2,0.58)
     (4.4,0.58)
     (4.6,0.58)
     (4.8,0.58)
     (5,0.58)

     (2.1,0.585)

     (2.0,0.59)
     (2.1,0.59)

     (2.0,0.6)
     (2.1,0.6)
     (2.2,0.6)
     (2.3,0.6)
     (2.4,0.6)
     (2.5,0.6)
     (2.6,0.6)
     (2.8,0.6)
     (3,0.6)
     (3.2,0.6)
     (3.4,0.6)
     (3.6,0.6)
     (3.8,0.6)
     (4,0.6)
     (4.2,0.6)
     (4.4,0.6)
     (4.6,0.6)
     (4.8,0.6)
     (5,0.6)

     (2.0,0.62)
     (2.1,0.62)
     (2.2,0.62)
     (2.3,0.62)
     (2.4,0.62)
     (2.5,0.62)
     (2.6,0.62)
     (2.8,0.62)
     (3,0.62)
     (3.2,0.62)
     (3.4,0.62)
     (3.6,0.62)
     (3.8,0.62)
     (4,0.62)
     (4.2,0.62)
     (4.4,0.62)
     (4.6,0.62)
     (4.8,0.62)
     (5,0.62)

     };

    \node at (axis cs:3.5,0.51) {\large \textbf{Isotropic}};    
    \node at (axis cs:3.5,0.59) {\large \textbf{Nematic}};

    \node[fill=white,inner sep=1pt,draw=black] at (axis cs:1.8,0.474) [anchor=north west] {compression/decompression $\updownarrow$};    
    \node[fill=white,inner sep=1pt,draw=black] at (axis cs:6.3,0.61) [anchor=south east] {\Large \textbf{HPR model}};
     
    \end{axis}
\end{tikzpicture}
\caption{Top: Phase diagram of hard PHGO pear-shaped particles with $k=3.0$ obtained by compression (from isotropic) and decompression at fixed tapering parameter $k_{\theta}$ for systems of $3040$ particles in a cubic simulation box. Grey regions between the isotropic and ordered phases indicate parameter values for which phase hysteresis is observed between compression and decompression sequences. The phase diagram is adopted from \zit{SEMCS-T2017}. Bottom: Phase diagram of hard HPR particles with $k=3.0$ obtained by compression (from isotropic) and decompression at fixed tapering parameter $k_{\theta}$ for systems of $400$ and $1600$ particles in a cubic simulation box. Grey shaded regions indicate configurations which showcase a high degree of local orientational order and basic features, which could lead to bilayer formations according to their pair-correlation functions (see \fig{fig:ori_hard}). However, this should not be seen as a separate phase from the isotropic state. The schematics above both graphs indicate the cross-sectional shape of the particles associated with each $k_{\theta}$ value.}
\label{fig:phase_diagram}
\end{figure*}

To highlight the sensitivity of the special collective behaviour of PHGO pears in terms of particle shape, the phase diagram of the PHGO pear-shaped particle model, which has been obtained in \cite{SEMCS-T2017}, is revisited and put into perspective in the following. In this previous paper a complete phase diagram of PHGO particles with aspect ratio $k=3$ is calculated (see also the recreated phase diagram in \fig{fig:phase_diagram}). Depending on the tapering parameter, the phase diagram can be separated into three regimes. Two parts, containing pears with high ($k_\theta<2.3$) and intermediate tapering ($2.3<k_\theta<4.5$), are characterised by the formation of bilayer-phases, namely the bilayer smectic and the gyroid configuration. The third fraction ($k_\theta>4.5$) of the phase diagram involves nearly ellipsoidal particles which generate monolayer states like nematic and monolayer smectic.\\

 \begin{figure*}[t!]
\centering
\input{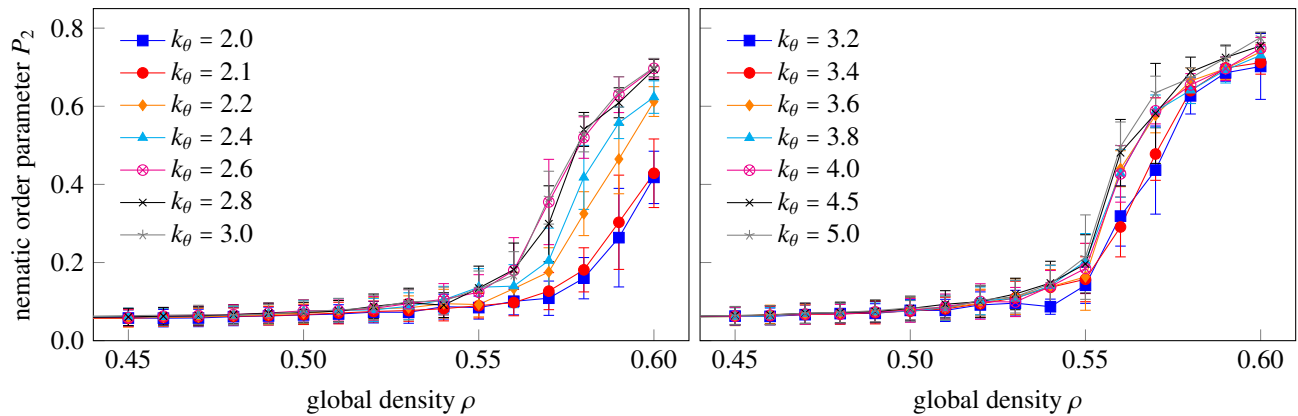}
\caption{The nematic order parameter $P_2$ during the compression of HPR particle systems with $N=400$ for different tapering parameters $k_{\theta}$.}
\label{fig:nemOrderHard}
\end{figure*}

\subsection{Phase behaviour of hard pears of revolution (HPR)}

The slight shape change of the pear particles are realised by changing the model to describe pear particle interactions from the PHGO to the HPR representation. The calculated phase diagram is based on $NVT$ Monte Carlo simulations with $N=400$ and $N=1600$ monodisperse HPR particles interacting via a hard-core potential. The boundary conditions of the cuboidal simulation box are set as periodic in all three directions. The tapering parameter $k_{\theta}$ lies between $2.0$ and $5.0$ which corresponds to tapering angles between $28.1^{\circ}$ and $11.4^{\circ}$. The MC translation step and the rotation step are initially set as $\Delta_{q,\text{max}} = 0.015\sigma_w$ and $\Delta_{u,\text{max}} = 0.015\sigma_w$ \footnote{The parameter $\sigma_w$ indicates the width of the pear-shaped particles.}, respectively, but have been adjusted in an equilibration phase to maintain acceptance rates of roughly 50\% for the displacement attempts.\\

Every simulation starts from an initially crystalline arrangement of particles at very low density ($\rho_g=0.1$), which is then compressed to the global density $\rho_g=0.44$ where all systems are obtained in the isotropic phase. Subsequently, the systems are slowly compressed further (see symbols in \fig{fig:phase_diagram}). For each data point of the sequence, the assembly is equilibrated for $2{\cdot}10^6$ MC steps and afterwards analysed for $1.8{\cdot}10^7$ step, where snapshots are taken after every 10000th step. At very high densities $\rho_g=0.63$, the mean squared displacement of the individual pears indicates trapped particles. Those particles hardly diffuse within the simulation box during simulation runs. This could be an indicator of a solid state. However, our simple Metropolis MC method is not sufficient to access this region reliably. Thus, solid phases are not drawn in the phase diagram. Afterwards, expansion sequences are performed in an equivalent, but reverse, manner from each $\rho_g=0.63$ state. The resultant phase diagram is shown in \fig{fig:phase_diagram}.\\
 
Already at first sight, the HPR phase diagram differs starkly from the phase diagram of PHGO particles. It becomes apparent that the remarkable division into three different regimes in terms of shape is absent. Independent of tapering all particles feature a similar phase behaviour. For low densities, the particles adopt the expected isotropic phase. However, during the compression, the pear-shaped particles begin to globally align with the director of the system and eventually transition into a nematic state (see nematic order parameter in \fig{fig:nemOrderHard}).\\

\begin{figure*}[t!]
\centering
\input{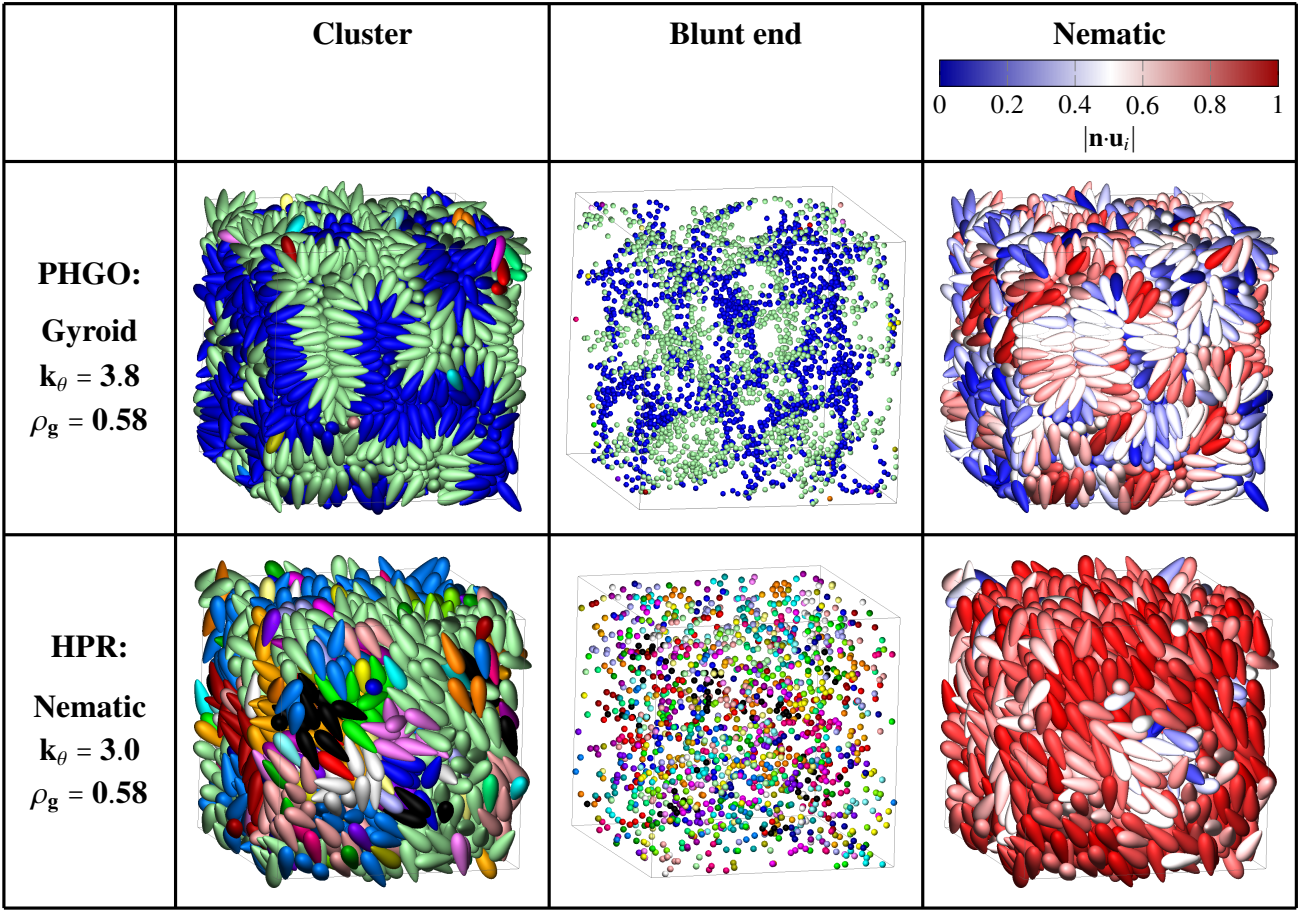}
 \caption{Representative configurations of $3040$ PHGO pear-shaped particles in the gyroid phase (first row: $k=3$, $k_{\theta}=3.8$, $\rho_g=0.60$) and $1600$ HPR particles forming the nematic phase (second row: $k=3$, $k_{\theta}=3.0$, $\rho_g=0.58$). The structures are illustrated in the cluster representation (first column) and the blunt end representation (second column) where the colors indicate the cluster affiliation. In the third column the particles are additionally colored according to their relative orientation to the director $\mathbf{n}$.}
\label{fig:phasesHard}
\end{figure*}

Also at direct visual comparison between the HPR and PHGO assemblies the major distinctions become apparent (see characteristic configurations pictured in \fig{fig:phasesHard}). Next to the absence of gyroid phases and of the global alignment into one preferred directions, the HPR particles even lack of any indications of bilayer formation. Neither do they display interdigitated zig-zag patterns of anti-parallelly aligned pears, nor is it feasible to detect layers or channel domains via distance clustering of their blunt ends for any given tapering parameter. By contrast the influence of the tapering parameter $k_{\theta}$ is manifested in a shift of the transition density from the isotropic to the nematic phase. A greater head-tail asymmetry of the pear shape induces destabilisation of the nematic order such that the transition occurs for larger densities. Also note that the hysteresis effects are marginal compared to those observed in the process of constructing \fig{fig:phase_diagram}. Consequently, the hysteresis is not drawn in this phase diagram. Moreover, the transition line coincides with previous observations of the isotropic-nematic transition for prolate ellipsoids with $k=3$ and $k_{\theta}{\rightarrow}\infty$ ($\rho_{in}=0.541$ \cite{FMMcT1984,O2012}). As the nematic phase arches over all values of $k_{\theta}$ it becomes evident that HPR pears seem to be unable to form bilayer-structures via self-assembly.\\

The computational complexity of the overlap calculations for HPR imply that our results are based on fewer and shorter simulation runs. While the question of equilibration is a more persistent one than for PHGO, there are clear indications that the HPR behaviour described above is a close representative of the equilibrium behaviour: Firstly, we have been unsuccessful in obtaining an equilibrated bilayer configuration even when the HPR systems are initially prepared as an artificial smectic or gyroid arrangement. Here the pre-constructed structures destabilise and transition into nematic configurations upon equilibration. Secondly, during our simulations the HPR pears hardly show any sign of precursors of bilayer formation. This, however, is a typical initial step in the isotropic phase of PHGO particles before entering the bilayer states \cite{SEMCS-T2017}. The precursors appear as small randomly oriented clusters which are unjoined such that they do not form long-ranged structures. Only HPR particles within the grey area in \fig{fig:phase_diagram} hint towards some of the characteristics of such bilayer precursors, which is discussed in more detail below.\\

\section{Pair correlation functions}
\label{sec:Pair}

Overall, we can draw the conclusion that the small differences between the PHGO and HPR model have major repercussions on the pears' ability to collectively form bilayer phases. To give an explanation for the drastic change in phase behaviour, we investigate the local surrounding of the different phases by calculating the lateral $g^{\perp}$ and longitudinal $g^{\parallel}$ pair-correlation functions. As the local behaviour is intimately linked with global phase behaviour, this analysis, next to our studies on the depletion behaviour of the two pear-shaped particle models in part 2 \cite{SMCS-T2020_2}, sheds light on the propensity of PHGO articles to form gyroid structures from a microscopic point of view. Here we concentrate not only on the density distribution in lateral and longitudinal direction of the pears, but also the polar and nematic weighted correlation functions. Before we apply these tools to the PHGO and HPR systems, however, we first describe the definition of $g(r)$ in detail, as a basis for our extended definition of $g^{\perp}$ and $g^{\parallel}$ below.\\

\subsection{Technical definition of pair correlation functions}
\label{sec:paircorrel}

\begin{figure*}[t]
\centering
\input{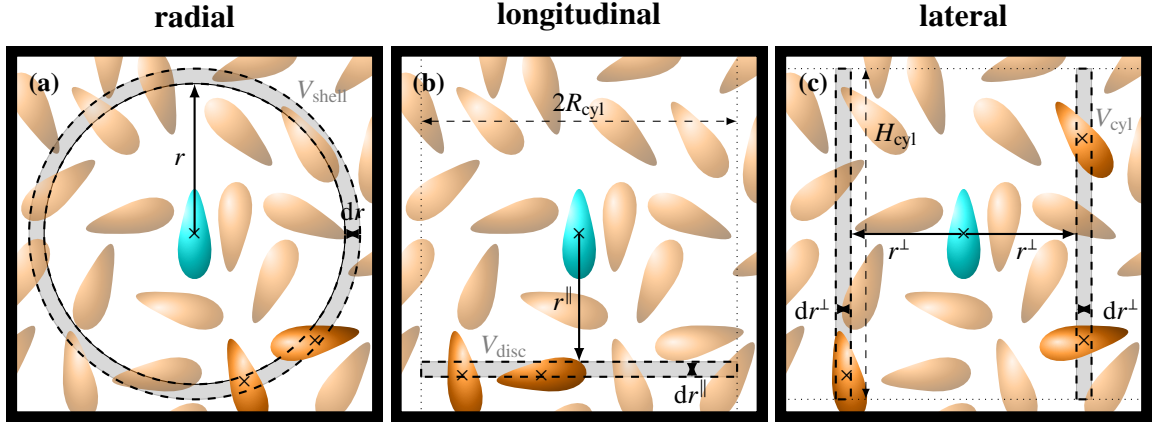}
\caption{Schematics of the radial (a), longitudinal (b) and lateral distribution function (c). The figures show cross sections through the sampling space. The gray areas represent shells which bin the space around the center pear-shaped particle and are used to create the corresponding histogram. The shells are spherical (a), discal (b) and cylindrical (c).}
\label{fig:DistributionFuction}
\end{figure*}

One of the best established observables to characterise the translational order of particle systems are the \textit{pair correlation function} $g(r)$, also known as the \textit{radial distribution function}. The radial distribution function represents the probability, given that particle $i$ is placed at the origin, to find another molecule $j$ at a radial distance $r$. Thus $g(r)$ bears valuable information about the positional correlations between the particles. Based on the number density distribution function 
the radial distribution function is written as
\begin{equation}
 \label{eq:RadialDistributionFunction}
g(r)=\frac{1}{N\rho_N}\left\langle\sum_i\sum_{j\neq i}\delta(r-r_{ij})\right\rangle
\end{equation}
with the global number density
\begin{equation}
 \label{eq:NumberDensity}
\rho_N=\frac{N}{V}.
\end{equation}
To calculate $g(r)$ numerically in our simulations, \eq{eq:RadialDistributionFunction} has to be discretised and rewritten. Based on the definition of $g(r)$, the mean number of particles $\delta N(r)$ found within a small distance interval $[r,r+\delta r]$ from another particle is given by
\begin{equation}
 \label{eq:RadialDistributionFunctionNumber}
\delta N(r)=\rho_Ng(r)V_{\text{shell}}(r)
\end{equation}
with $V_{\text{shell}}(r)$ being the volume of the thin spherical shell of thickness $\delta r$ whose inner boundary is a sphere of radius $r$. By approximating $V_{\text{shell}}(r)=V_{\text{sph}}(r+\delta r)-V_{\text{sph}}(r)\approx 4\pi r^2 \delta r + \mathcal{O}(\delta r^2)$ and rearranging \eq{eq:RadialDistributionFunctionNumber}, we obtain
\begin{equation}
 \label{eq:RadialDistributionFunctionHistogram}
g(r)=\frac{1}{\rho_N}\frac{\delta N(r)}{4\pi r^2\delta r}.
\end{equation}
This can be interpreted as a formula to generate the radial distribution function by a normalised histogram. The histogram is computed by counting all pair separations, corresponding to the domain $m\delta r < r_{ij} < (m+1)\delta r$ and normalize them according to \eq{eq:RadialDistributionFunctionHistogram}. Note that the ``normalisation'' factor in this case indicates that $g(r)$ converges towards 1 for large distances: $\lim_{r\rightarrow\infty}g(r)=1$. This indicates that a pair of particles at large distance from one another is uncorrelated. Additionally, to prevent boundary effects only pairs with $r_{ij}<\frac{L}{2}$ are considered in calculating $g(r)$. The concept is pictured in \fig{fig:DistributionFuction}a.\\ 

In the analysis of liquid crystals it is often advantageous not to determine the radial distribution as described above, but to separate the distance between two molecules into a longitudinal and a lateral part, particularly for smectic phases. Due to their anisotropic features, the order parallel to the director is different from the order perpendicular to the director. By calculating $g^{\parallel}(\mathbf{n}\cdot\mathbf{r})$ and $g^{\perp}(\sqrt{r^2-(\mathbf{n}\cdot\mathbf{r})^2})$ the information is separated for the two directions. The former characterises the smectic layering of the system, whereas the latter is a measure of translational order within the layers. However, this approach has the disadvantage that global orientational order is needed. Lipid systems adopting a bicontinuous surface geometry, exhibit no overall global orientational order as they form pronouncedly curved bilayers. Nevertheless, locally neighbouring lipids are clearly orientationally correlated such that a lateral and longitudinal distribution function on a local scale seems to be more effective. Thus, we replace the director with the orientation of the liquid crystal at the origin $\mathbf{u}_i$. In this way, we can guarantee to detect both curved bilayer ordering but also smectic layering as $\mathbf{u}_i\approx \mathbf{n}$ \footnote{This only applies to the smectic-A phase. For other smectic phases it is still more convenient to use the director as a reference.}. The longitudinal and lateral distance are defined by $r^{\parallel}=\mathbf{u}_i\cdot\mathbf{r}$ and $r^{\perp}=\sqrt{r^2-r^{\parallel 2}}$, respectively. Note here, that $r^{\parallel}$ can become negative. For pear-shaped particles, positive longitudinal distances correspond to a distance in the direction of the thin narrow end while negative distances have to be assigned to particles which are placed in the direction of the thick blunt end.\\

To compute the longitudinal distribution function $g^{\parallel}(r^{\parallel})$ and lateral distribution function $g^{\perp}(r^{\perp})$ we use a similar histogram approach like in \mbox{\eq{eq:RadialDistributionFunctionHistogram}.} For simplifying the normalisation of the histograms they are calculated within a cylinder. This implies that only particles which lie within a cylinder with radius $R_{\text{cyl}}$ and height $H_{\text{cyl}}$ centered at the position of particle $i$ are considered. The cylinder, furthermore, shares the same rotational symmetry axis as the very particle $i$ (see \fig{fig:DistributionFuction}b). The dimensions of the encapsulating cylinder have to be chosen such that the periodic boundaries of the simulation box are not trespassed
\begin{equation}
 \label{eq:CyliderDimensions}
 \begin{aligned}
    H_{\text{cyl}} &< L\sin{\alpha}\\
    R_{\text{cyl}} &< \frac{L}{2}\sin{\alpha}.
 \end{aligned}
\end{equation}
Here, $\alpha$ encodes the aspect ratio of the cylinder $\tan\alpha$. The probability to find a particle at longitudinal distance $r^{\parallel}$ within a circular disk of thickness $\delta r^{\parallel}$ and volume $V_{\text{disc}}=\pi R_{\text{cyl}}^2\delta r^{\parallel}$ bounded by the cylinder is given by
\begin{equation}
 \label{eq:LongitudinalDistributionFunctionHistogram}
g^{\parallel}(r^{\parallel})=\frac{1}{\rho_N}\frac{\delta N^{\parallel}(r^{\parallel})}{\pi R_{\text{cyl}}^2\delta r^{\parallel}}.
\end{equation}
$\delta N^{\parallel}(r^{\parallel})$ is the mean number of particles within the disc. Analogously, probability to find a particle at lateral distance $r^{\perp}$ within a cylindrical shell of thickness $\delta r^{\perp}$ and volume $V_{\text{disc}}\approx 2\pi r\delta r^{\parallel}H_{\text{cyl}}$ is defined as 
\begin{equation}
 \label{eq:LateralDistributionFunctionHistogram}
g^{\perp}(r^{\perp})=\frac{1}{\rho_N}\frac{\delta N^{\perp}(r^{\perp})}{2\pi H_{\text{cyl}}r^{\perp}\delta r^{\perp}}.
\end{equation}
Here $\delta N^{\perp}(r^{\perp})$ is the mean number of particles within the cylindrical shell. The notion of both distribution functions is depicted in \fig{fig:DistributionFuction}b+c.\\

The different distribution functions provide the possibility to study the local orientational ordering in a much more detailed way as well. Here, the number density in \eq{eq:RadialDistributionFunction} can be weighted by a factor which includes the relative orientations of the pear particles. With this take on $g(r)$ we can define a polar radial distribution function $g_{P1}$ weighted by the first Legendre polynomial $P_1(\mathbf{u}_i\cdot\mathbf{u}_j)=\cos(\mathbf{u}_i\cdot\mathbf{u}_j)$
\begin{equation}
 \label{eq:PolarRadialDistributionFunction}
g_{P_1}(r)=\frac{1}{N\delta N(r)}\left\langle\sum_i\sum_{j\neq i}\cos(\mathbf{u}_i\cdot\mathbf{u}_j)\delta(r-r_{ij})\right\rangle.
\end{equation}
For the nematic radial distribution function $g_{P2}$ the second Legendre polynom $P_2(\mathbf{u}_i{\cdot}\mathbf{u}_j)=\frac{1}{2}(3\cos^2(\mathbf{u}_i\cdot\mathbf{u}_j)-1)$ is used as weighting factor, such that
\begin{equation}
 \label{eq:NematicRadialDistributionFunction}
g_{P_2}(r)=\frac{1}{N\delta N(r)}\left\langle\sum_i\sum_{j\neq i}\frac{1}{2}(3\cos^2(\mathbf{u}_i\cdot\mathbf{u}_j)-1)\delta(r-r_{ij})\right\rangle.
\end{equation}
Both the polar and nematic distribution function are scaled by the mean number of particles at distance $r$ to easier relate the values to polar and nematic order parameters. This means that $g_{P_1}(r)$ and $g_{P_2}(r)$ determine how strongly two particles separated by a distance $r$ are orientationally correlated. However, the functions do not contain information about the likeliness of such configurations occurring. In a similar vein also lateral and longitudinal variants of the distributions are defined.
\begin{figure*}[t!]
\centering
\input{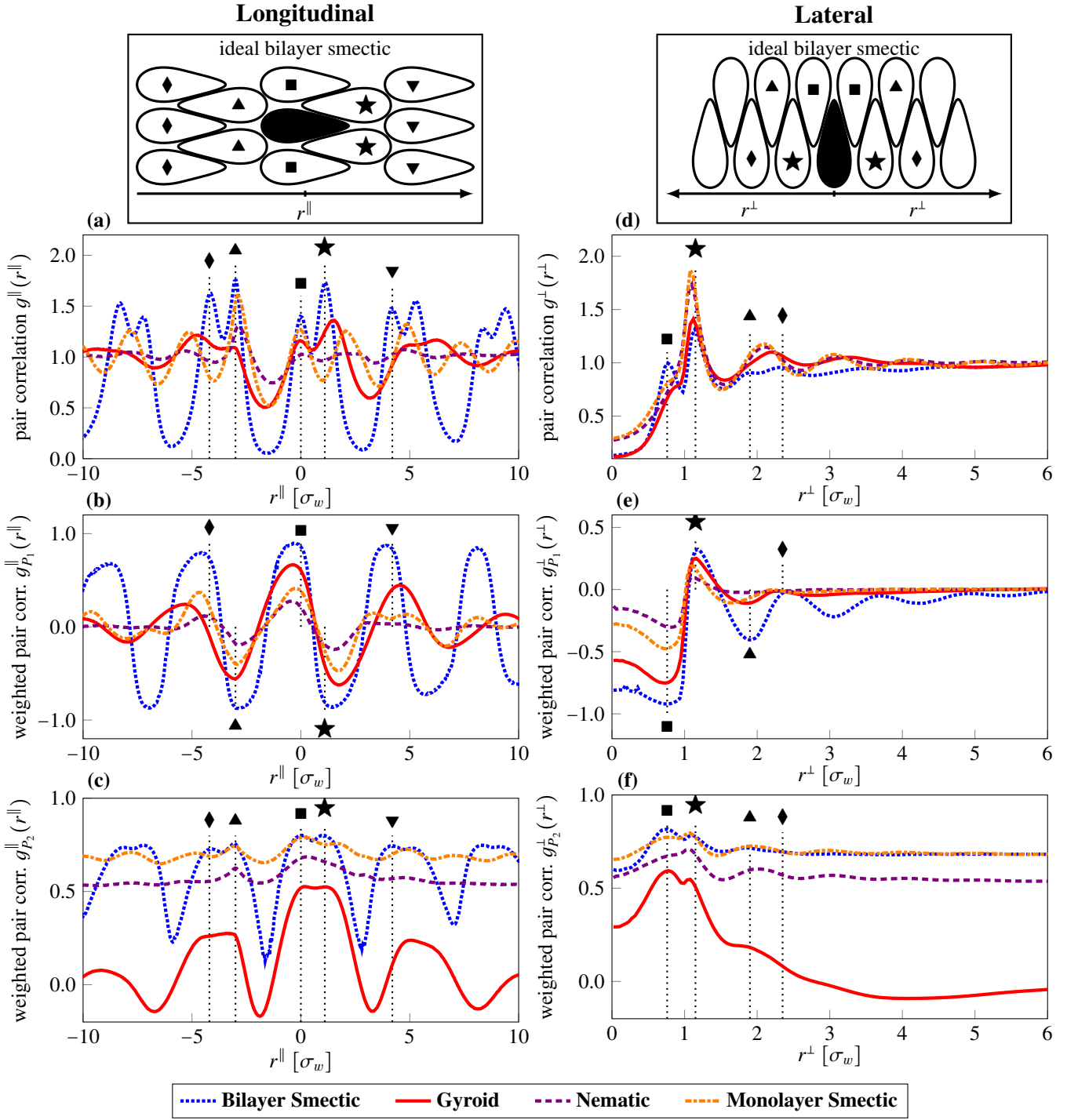}
\caption{The longitudinal pair-correlation function $g^{\parallel}(r^{\parallel})$ (left column) and the lateral pair-correlation function $g^{\perp}(r^{\perp})$ (right column) of the smectic bilayer ($k_{\theta}=2.2$,$\rho_g=0.57$), the gyroid ($k_{\theta}=3.8$,$\rho_g=0.56$), the nematic ($k_{\theta}=5.4$,$\rho_g=0.56$) and the smectic monolayer phase ($k_{\theta}=5.4$,$\rho_g=0.585$). The pair-correlation functions are additionally weighted by the polar order parameter $P_1$ (second row) and the nematic order parameter $P_2$ (third row).}
\label{fig:ori_go}
\end{figure*}

\subsection{Pair correlation functions of PHGO systems}

The lateral and longitudinal pair correlation functions are first applied to various PHGO systems which represent the different phases in the phase diagram shown in \fig{fig:phase_diagram}. The local properties to form bilayers have a clear signature in the form of the different longitudinal pair-correlation functions $g^{\parallel}(z)$ of PHGO particles (see \fig{fig:ori_go} left). In case of the smectic bilayer phase, all three plots (a-c) indicate multiple distinct peaks suggesting both long ranged transitional, polar and nematic order in the longitudinal direction but also a piling of multiple sheets of pear-shaped particles. Moreover, the bifurcation of peaks in \fig{fig:ori_go}a, for instance the pair of peaks indicated by $\blacksquare$ and $\bigstar$, implies an organisation into stacks of interdigitated bilayers rather than monolayers. Here, the arrangement into parallel leaflets ($\blacksquare$, $\blacklozenge$, $\blacktriangledown$), where the polar order parameter $P_1$ locally exhibits positive values, and antiparallel leaflets of the bilayers ($\bigstar$, $\blacktriangle$), where $P_1$ changes sign, can be identified. This propensity to obtain local polar order is also observed in pear-sphere-mixtures dominated by small hard spheres, where the PHGO particles align due to depletion attractions (see part 2 of this series \cite{SMCS-T2020_2}). The leaflets are also affirmed by the $g_{P_2}^{\parallel}(z)$ profile of this phase in the form of small dips at each maximum. Also the lateral pair-correlations indicate the smectic bilayer phase (see \fig{fig:ori_go} right). Firstly, the weighted functions show that the particles are aligned for large lateral distances suggesting that the layers are flat. Secondly, a small peak ($\blacksquare$) before the main peak is observable in \fig{fig:ori_go}d+f, which can be assigned to the immediate antiparallel and parallel neighbours of the reference pears in the same bilayer, respectively.

Analogously the pair correlation functions belonging to gyroid forming PHGO particle systems prove that single particles arrange within interdigitating curved bilayers. The characteristics of the distance distributions are locally similar to those observed in the flat bilayer-smectic phase of strongly tapered pears. The bifurcation of peaks (a) and the clear bump at the location of the secondary minor maximum for small $r^{\perp}$ in the bilayer smectic phase (d) coincide with the architecture of interdigitated bilayers. Yet, both of these plots also point to considerable differences on a larger length scale. The correlations are less distinct and diminish faster in the longitudinal and lateral direction which can be explained by the inherent curvature of the minimal surface structure. The influence of the warped bilayers is reflected even more in the characteristics of the weighted pair correlation functions. Firstly, the polar order vanishes in (b+e) for large distances and is less periodic. Secondly the nematic order in (c) \mbox{oscillates} around $0$ and, like the plot in (f), eventually approaches this very value for $r^{\parallel}\rightarrow\infty$. This means that the stacks of bilayers do not lie parallel to each other anymore and also that largely separated particles within the same leaflet are likely to be differently oriented.\\

Also the pair-correlation functions of the nematic and monolayer smectic give valuable information about the importance of the mentioned signatures of the different $g(r)$s for bilayer assembly. Although both translational and orientational order is still present, the correlations are weaker than for bilayer arrangements. Furthermore, the plots not only differ quantitatively but also qualitatively. On the one hand, the division into two maxima per peak for $g^{\parallel}(r^{\parallel})$ in \fig{fig:ori_go}a vanishes. On the other hand, the small secondary peak which was contributed to the opposite leaflet of a bilayer also disappears for small $r^{\perp}$ in $g^{\perp}(r^{\perp})$ (see $\blacksquare$ in \fig{fig:ori_go}d). Both of these phenomena can be explained by the lack of inversion asymmetry. In this regime, the particles are not tapered enough to interdigitate into a neighbouring sheet and rather form a separate monolayer. Moreover, the weak taper causes the polarity within a sheet to be less pronounced (indicated by the overall small peaks in the $P_1$ profiles) as in the bilayer smectic phase, such that antiparallel particles can be found within the same leaflet more often (high peak at $\bigstar$ in \fig{fig:ori_go}d). This also causes the profile of the nematic and monolayer smectic phases in \fig{fig:ori_go}c to be more homogeneous at a high mean nematic value.\\

\subsection{Pair correlation functions of HPR systems}

\begin{figure*}[t!]
\centering
\input{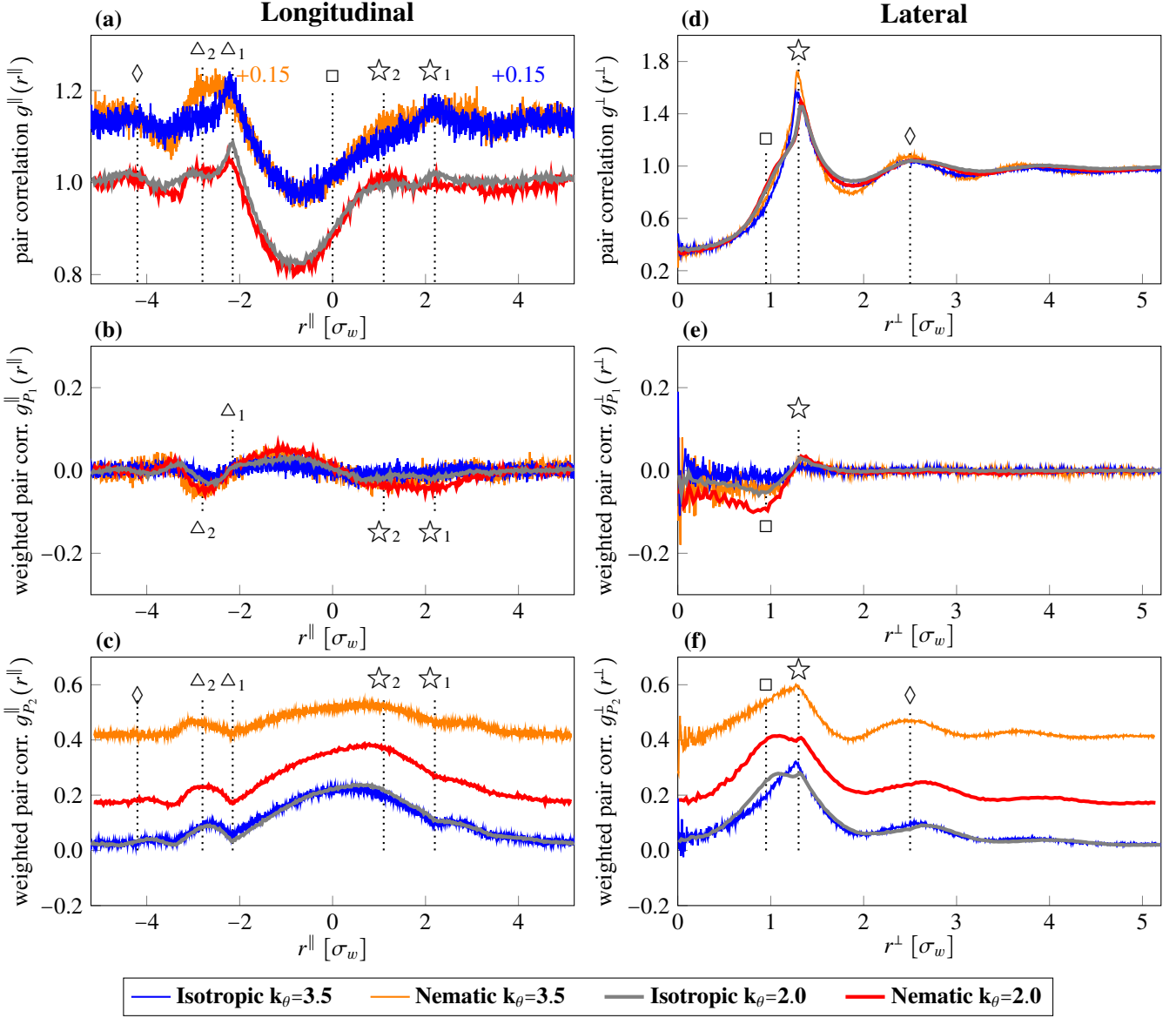}
\caption{The longitudinal pair-correlation function $g^{\parallel}(r^{\parallel})$ (left column) and the lateral pair-correlation function $g^{\perp}(r^{\perp})$ (right column) of the isotropic ($k_{\theta}=2.0$:$\rho_g=0.58$ and $k_{\theta}=3.5$:$\rho_g=0.55$) and nematic ($k_{\theta}=2.0$:$\rho_g=0.6$ and $k_{\theta}=3.5$:$\rho_g=0.58$) in systems of $N=400$ HPR particle. The pair-correlation functions are additionally weighted by the polar order parameter $P_1$ (second row) and the nematic order parameter $P_2$ (third row).}
\label{fig:ori_hard}
\end{figure*}

Based on these observations gained from the PHGO particles, we can deduce the lack of bilayer phases in the HPR phase diagram, by an analysis of these phases' local behaviour. The profiles of the pair correlation functions in the nematic and the isotropic phase close to the transition line (see \fig{fig:ori_hard}) exhibit both similarities and differences to the liquid crystal phases of the PHGO pear systems in \fig{fig:ori_go}. The lateral pair-correlation functions $g^{\perp}(r^{\perp})$ of the nematic phases of both pear models, for example, produce similar plots, also comparable to the monolayer smectic of the PHGO model. The characteristic minor peak before the first major peak (see $\square$ in \fig{fig:ori_hard}d), however, which have been attributed to interdigitating bilayer arrangements, is not present. Only for pears close to $k_{\theta}=2.0$ this peak is implied by a bump. Also the profiles of $g_{P_2}^{\perp}(r^{\perp})$ are akin (even if the alignment is not as strong) to the not-bilayer forming liquid crystal phases of the weakly tapered PHGO pears. The most significant difference in terms of lateral correlation, however, is in the polarity of the neighbouring particles in \fig{fig:ori_hard}e. For HPR pears the nearest neighbours show basically no preference of parallel or anti-parallel orientation. The high degree of local polar order for PHGO pears is at best vaguely reflected and largest for $k_{\theta}<2.5$.\\

The plots of the longitudinal pair correlations $g^{\parallel}(r^{\parallel})$ shown in \fig{fig:ori_hard} left, however, also indicate why the particles are not arranged within a bilayer formation and rather create nematic phases. The most noticeable one is the missing peak ($\square$ in \fig{fig:ori_hard}a) at $r^{\parallel}=0$ in the nematic and monolayer smectic phase. This signifies that this particular correlation is crucial for the formation of bilayer phases as it corresponds to particles sitting side by side to another. All other peaks ($\largestar$,$\triangle$,$\lozenge$) can be attributed to their counterparts in the $g^{\parallel}(r^{\parallel})$-signature of the nematic/smectic phases of the PHGO pears, but seem to be closer together. Furthermore, the weighted functions indicate that the reference pears barely influence the polar preference of their neighbour's orientation, not even longitudinal direction. On a similar note, the local nematic order indicated by the minor peaks, even though obviously present, is not as pronounced and long ranged in this model, not to mention the double peaks, which can be observed for all liquid crystal phases in \fig{fig:ori_go}, but are not noticeable here. \\

Despite these distinctions, similarities can be determined as well. For once, the pears tend to aggregate preferentially at the blunt ends ($r^{\parallel}<0$) rather than the pointy end ($r^{\parallel}>0$) of other particles. This leads to the assumption that in principle the mechanism which brings the pears together with their blunt ends to form clusters also exists in the HPR model. Unfortunately, the impact of this mechanism is not strong enough to indeed induce the self-assembly of bigger clusters (see cluster representation in \fig{fig:phasesHard}). More intriguing, however, is the observation that for highly tapered particles $k_{\theta}<2.5$ the peaks of $g^{\parallel}(r^{\parallel})$ ($\largestar_1$,$\largestar_2$ and $\triangle_1$,$\triangle_2$) and $g_{P_2}^{\perp}(r^{\perp})$ ($\square$,$\largestar$) widen considerably or even split into two. This can be already observed in the isotropic phase close to the phase transition. The area within the system which showcases these indications of bifurcation is shaded in the phase diagram. Thus, some of the basic conditions for bilayer formation are also met at least for highly tapered HPR particles. Nevertheless, without additional features to the contact function, those effects are too weak to produce a more complex phase behaviour than nematic.\\

In this paper, we focused exclusively on pear-shaped particles with a specific aspect ratio of $k=3$. While possible, it is unlikely that a different choice of $k$ for the HPR would have yielded a different phase behaviour, for the following reasons. Firstly, by increasing the aspect ratio, the maximum adjustable taper of convex pear-shaped particle decreases. As we have shown that higher taper implies higher local order, we can rule out the existence of the gyroid phase in HPR systems for $k\geq 3$. Secondly, less elongated hard particles usually lose their ability to create global orientational order (rule of thumb $k<2.75$ \cite{VF1990,FMMcT1984}) and form isotropic configurations instead. Therefore, the window of aspect ratios, which comes into consideration, seems too small to increase the local polar order in \fig{fig:ori_hard}b+e to values, which are needed to achieve bilayering comparable to PHGO systems. 

\section{Conclusion and outlook}

The overarching theme of this paper concerned the stability of the gyroid phase with respect to particle shape, particularly the difference in phase behaviour between HPR and PHGO particles. It hence fits closely with the broader topic of how self-assembly (in particular in hard core systems) is sensitive to the details of the particle shape \cite{BST2010,ZFvdLG2011,NGdeGvRD2012,RSAPPCDSI2015,DEG2011,DEG2012,GdeGvRD2013,DvANG2017,KCEG2018,DD2016,MD2010,WDvAG2019}. In particular, we compared two hard pear-shaped particle models on the microscopic scale and their abilities to form the double gyroid spontaneously globally. One is the pear hard Gaussian overlap (PHGO) particle, which closely approximates a pear-shape but also features self-non-additive properties. The other model represents the exact pear shape perfectly and is called hard pear of revolution (HPR) model.\\

Therefore, we revisited the phase behaviour of PHGO particles and additionally generated a phase diagram based on particles interacting according to strict hard-core HPR interactions. In contrast to the rich phase diagram of PHGO particles containing nematic and monolayer smectic, but also both bilayer smectic and bilayer gyroid structures, we observed in the HPR systems only a rudimentary phase behaviour. More precisely, the HPR systems form nematic liquid crystal phases for all particle shapes analysed (i.e. all $k_\theta$), where more highly tapered particles visibly destabilise the nematic order and push the transition to higher densities. However, both the gyroid and the bilayer smectic phase, characteristic for the phase behaviour of PHGO particles, vanish.\\ 

According to these observations the small differences in the contact function between the PHGO and HPR model, which can easily, but mistakenly, be considered negligible, have a major impact on the self-assembly of pear-shaped particles. Even though most features of a pear (like aspect ratio and tapering parameter) are present in both models, the PHGO particles have to offer additional morphological properties, to which the stability of the gyroid phase is ascribed. This is also supported by the fact that only the nematic phase is obtained which also have been found for PHGO pears with small tapering angles. In this regime of large $k_{\theta}$ the two pear models differ the least in terms of contact functions. Hence, their collective behaviours are very similar. All these results lead to the assumption that the formation of bilayer structures, including the double gyroid phase, is due to the special orientation dependency of the PHGO contact function. Especially the self-non-additive features in reference to the pear shape seem to magnify the spontaneous placement of pears side to side. This mechanism would naturally lead to sheets, which then interdigitate due to the pointy ends of the individual particles. Not only the HPR model and our depletion studies in part 2 \cite{SMCS-T2020_2} hint towards the validity of this hypothesis, also other models which lack self-non-additive features but look similar to pears are known to fail assembling into bilayer configuration. Neither hard multisphere particles, like snowman \cite{DMD2012} or asymmetric dumbbell particles \cite{MDD2013}, nor conical colloids \cite{vdH2016} show any propensity to form the gyroid.\\

Despite the differences in phase behaviour, the self-assembly of some HPR particles with small $k_{\theta}$ close to the phase transition showcases also interesting properties, which were attributed as necessary precursors to the formation of bilayers. Therefore, it is conceivable that the HPR particles might be able to form similar phases like the PHGO pears, if we, for instance, add suitable changes to the pear-shape or introduce non-additivity to the HPR contact function. These particle modifications also have the potential to be utilised as a regulating mechanism to control the coupling strength between the blunt ends. This might allow us to create a model for pear-shaped particles, based on those indicated by the grey-striped area in \fig{fig:phase_diagram}, with an intermediate degree of blunt end aggregation. A first attempt to conceptualise such a pear-shaped particle model is made in part 2 of this series \cite{SMCS-T2020_2}. In general, these particles could potentially form phases with a short-range order, sufficient to display a bicontinuous network, but also displays with disorder over larger length scales. Those disordered cubic phases are known as L$_3$ sponge phases \cite{AWO1989} and are formed typically in lipid-water mixtures by swelling the cubic phases due to the presence of additives \cite{EARL-W1998,ESLE2002,CCPC2006,AT2007,WJW-HWFHKNOeYCFEN2008,L1994,BMGTJ2006,I-PA-CM2011,VWR-OKDNB2016}.\\

The formation of gyroid structures in pear-shaped PHGO particle systems remains a fascinating finding. This is particularly so because of the mechanism of creating a propensity for the formation of interdigitated ``smectic-like'' warped bilayers. While particle shape clearly plays a crucial role in this, this paper has highlighted the subtleties, namely that the effect vanishes for the additive hard pear HPR model. This, in turn, brings us back to the opening statement that the particle shape is a double-edged sword. Surely, the ``coarse'' (or first order) characterisation of the particles as pear-shaped is critical for the process. Yet, pear-shaped appearance is not sufficient to ensure the effect occurs, as the lack of the gyroid in the HPR phase diagram demonstrates. It appears as first-order shape characteristics are a necessary condition for some structure phase formation but not a sufficient criteria.\\   

As a closing note, we want to mention here that it is difficult to judge which of the two pear models represents the interactions of pear-shaped particles, which might be synthesised in the future, better. For example, it is well established that colloids in experimental systems are never truly hard and the interparticle potential always inherits some degree of softness \cite{PvM1986,PvM_SC1987,PWR2012,RPW2013}. Therefore, the potentials we used here -- both the PHGO \textit{and} the HPR potentials -- have to be considered as approximations of a real pear-shaped colloid. This becomes even more important as recent studies show that the introduction of already a small degree of softness can influence the stability of crystalline phases \cite{LaCADG2019}. Additionally, pear-shaped particles have not been synthesised yet. In principle, many different strategies to produce nanoparticles with aspherical shapes have been developed like methods via templates \cite{RMEEDDeS2005,LGJWECL2008,CAD2015}, particle swelling and phase separation \cite{KLW2006,KBEP2006,PFD2009}, seeded emulsion polymerisation \cite{PDLTB-LR2009,PI2015,FLSLLW2018,LVVGRSTDS2018}, controlled deformation of spherical colloids \cite{ZvWI2006,CKM2007,SK2012}, particle confinement \cite{YLGX2001} or lithography \cite{DPCHD2006,GFLH2006,LeGLGHD2015}. However, many of these techniques are still limited in either their customizability of the particle shape, rely on colloids as a basic shape or cannot be mass-produced easily. These difficulties seem to be exacerbated by the big contrast of the two phase diagrams in \fig{fig:phase_diagram}, which highlights that in both experiments and simulations even small nuances of the interaction profiles of molecules have to be taken into account to predict the right phase behaviour. Also the composite sphere method, where complexly shaped particles are modelled from multiple sphere constituents, are known to faces issues with inaccuracies due to the degraded smoothness of the particle surface \cite{K-ERWS2008,MKDN2010,HWK-ES2011}.

\begin{acknowledgments}
We thank Universities Australia and the German Academic Exchange Service (DAAD) for funds through a collaboration funding scheme, through the grant ``Absorption and confinement of complex fluids''. We also thank the DFG through the ME1361/11-2 grant and through the research group ``Geometry and Physics of Spatial Random Systems'' (GPSRS) for funding. We gratefully acknowledge Klaus Mecke's support and advice in useful discussions. P.W.A.S. acknowledges a Murdoch University Postgraduate Research Scholarship. G.E.S-T is grateful to the Food Science Department at the University of Copenhagen and the Physical Chemistry group at Lund University for their hospitality and to Copenhagen University, the Camurus Lipid Research Foundation and the Danish National Bank for enabling a sabbatical stay in Denmark and Sweden.

\end{acknowledgments}

\subsection*{Data availability}
The data that supports the findings of this study are available within the article. Data set lists are available from the corresponding authors upon reasonable request.

\bibliographystyle{unsrt}
\bibliography{reference}
\end{document}